# Enhancing Cutoff Energy of Solid High-Harmonic Generation from Bonding Length Perspective

Qing-Guo Fan[1,2], Kang Lai[1], Wen-hao Liu[1], Lin-Wang Wang[1,2,*], Jun-Wei Luo[1,2,**], Zhi Wang[1,†]

[1] State Key Laboratory of Semiconductor Physics and Chip Technologies, Institute of Semiconductors, Chinese Academy of Sciences, Beijing 100083, China

[2] Center of Material Science and Optoelectronics Engineering, University of Chinese Academy of Sciences, Beijing 100049, China

**Abstract**: High-harmonic generation (HHG) from solid state offers promising potential for attosecond optics with enhanced efficiency and compact configurations. However, Current implementations face critical limitations imposed by material damage thresholds, directly restricting spectral cutoff energies in nonperturbative regime. In this study, we control the cutoff energy through tailoring the bond length of materials, which is available by experimental strain. Employing real-time time-dependent density theory (rt-TDDFT) simulations, we find that the cutoff energy increases by nearly one third under a bond length compression of 7.5%. Our results reveal that it originates the band gap widening inducing the enhancement of interband cutoff energy, which is material-independent. This work provides novel theoretical insights for optimizing extreme ultraviolet sources, advancing potential applications in attosecond physics.

**Keywords**: High-harmonic generation, solid state, time-dependent density functional theory

**Corresponding authors:**
*     lwwang@semi.ac.cn
**    jwluo@semi.ac.cn
†     wangzhi@semi.ac.cn

## I. INTRODUCTION

HHG serves as the cornerstone of attosecond science, which enable gating the isolated attosecond pulses (IAPs) from broadband spectral[1–5]. These extremely ultrashort bursts have emerged as indispensable tools for probing electron and molecular dynamics in modern ultrafast time-resolved metrology[6–8].

Although initial investigations predominantly occurred in gas-state HHG[9–12], solid-state HHG has garnered significant research attention in recent years[13–23] owing to its enhanced efficiency and controllability stemming from dense periodicity[24]. More intriguingly, its fundamentally distinct physical mechanisms have enabled novel applications in probing material properties, such as bandstructure[25], lattice dynamics[26–28], topology[29,30], and atomic potentials[31]. Nevertheless, material damage thresholds impose stringent limitations on achievable cutoff energies, hindering practical extreme ultraviolet (XUV) source development[17,19].

Current research efforts to enhancing cutoff energy reveal unique scaling laws: solid HHG cutoff energy predominantly exhibits linear scaling with laser peak field strength[14,18,19,21,32–35], contrasting with the quadratic relationship governing gas state harmonics[12], and mixed linear-quadratic regimes occasionally emerge under specific conditions[36,37]. The wavelength dependence remains contentious, with some studies suggesting proportional relationships[14,34,35] while others posit wavelength independence[15,38].

Besides, one also considers material engineering approaches, including nanostructuring[39], metasurfaces[40,41], and doping[42–45], have demonstrated HHG modulation capabilities. Recent investigations into strain-engineered materials reveal that strain typically reduces HHG yield through bandgap widening[46–51]. This mechanism originates from bond-length modification: increased interatomic distances weaken orbital hybridization, diminish electron cloud overlap, reduce Coulomb repulsion, and consequently decrease the split between bonding and anti-bonding, which finally collectively suppressing electronic excitation[52,53]. However, existing research predominantly addresses yield reduction while overlooking strain-induced effects on cutoff energy parameters.

In this work, we employ ab initio calculation within time-dependent density functional theory framework to investigate the bond-length effects on the cutoff energy in solid-state HHG. With cubic bulk crystalline silicon and aluminum arsenide as prototypes, we find the cutoff energy could be improved by compressing bond length. Through wavelet time-frequency analysis, we further resolve the temporal emission mechanisms of harmonics generation.

## II. METHODS

### A. DFT simulation

All ab initio calculations are performed with the ab initio package PWMAT[54,55]. We utilize the optimized norm-conserving pseudopotentials Vanderbilt[56] and adiabatic Perdew-Burke-Ernzerhof (PBE) exchange-correlation functions. The expansion of the evolved wavefunction is based on plane-wave basis with a cutoff energy of 50 Ry, and an $12^3$ k-point mesh with Monkhorst-Pack grids to sample the irreducible Brillouin zones of primitive cell for Si and AlAs, respectively. Si (2s2 2p2) are considered as valence electrons, Al (2s2 2p6 3s2 3p1) and As (4s2 4p3) are considered as valence electrons in simulation. We modify the bond length like applying material isotropic strain. The relaxation of Si and AlAs primitive cell are chosen as $10^{-3} eV/Å$.

As is well known, DFT can't avoiding to underestimate the band gap of materials due to the limitations in accounting for self-interaction[57] and excitonic effects[58]. However, neglecting this aspect seldom affects our conclusions. Additionally, effects such as spin, dephasing caused by lattice and impurity scattering[15] and carrier lifetime are also disregarded. Furthermore, common nonlinear optical phenomena are not taken into account, like self-focusing and propagation effects. The balanced primitive cell of bulk silicon and cubic phase of AlAs with a lattice constant $a = 5.470 Å$ and $a = 5.657 Å$, respectively.

## B. TDDFT simulation

We employ the sine square envelope with full width at half maximum (FWHM) of 20fs and a center carrier wavelength of 3000 nm (corresponding to $0.60 eV$) without carrier-envelope phase. The laser of peak intensity is fixed at $1.0$ TW/cm$^2$ (in vacuum, ignoring surface reflection, below the material's damage threshold[59]), which is polarized along [111] ($\Gamma K$) direction. We use a dense kmesh grids of $28^3$ to get convergence current results.

## C. Calculating HHG and yield

We solve the time-dependent Kohn-Sham equations to obtain the Kohn-Sham orbital $\Psi_i(r,t)$ basing on wavefunction expansion by adiabatic eigenstates basis[55]. Here, $i$ denotes the band index. The laser potential is considered in velocity gauge:

$$i\frac{\partial}{\partial t}\Psi_i(r,t) = \left(\frac{1}{2}(-i\nabla + A^2) + v_{ext}(r,t) + v_{HF}[n(r,t)] + v_{xc}[n(r,t)]\right)\Psi_i(r,t) \quad (1)$$

where $v_{ext}$ is the electron-ion potential, $v_{HF}$ is the Hartree part of the Coulomb electron-electron interaction, $v_{xc}$ is the exchange-correlation potential, $A$ is the laser vector potential, which is simulated in dipole approximation[60]: $\mathbf{A}(\mathbf{r},t) \approx \mathbf{A}(t) = A_0 sin^2(t\pi/\delta)\sin(\omega_0 t)$, where $A_0$ is the peak potential, $\delta$ is the total elapsed time, $\omega_0$ is the carrier angular frequency, the electric field $\mathbf{E}$ can be derived from the relationship: $\mathbf{E}(t) = -\partial \mathbf{A}(t)/\partial t$. We can get the time-dependent electron spatial density by $n(\mathbf{r},t) = \sum_i |\Psi_i(r,t)|^2$. The microscopic current $\mathbf{j}(\mathbf{r},t)$ can be computed as:

$$j(r,t)_i = \frac{1}{2}(\langle\Psi_i(r,t)|-i\nabla + A|\Psi_i(r,t)\rangle) + c.c. \quad (2)$$

where $c.c.$ denotes the complex conjugate. The HHG spectrum is obtained from the

Fourier transform $(\mathcal{FT})$ of the first derivative of time-dependent current:

$$\text{HHG}(\omega) \propto \left| \mathcal{FT} \left\{ \frac{\partial}{\partial t} \int_\Omega d^3 \boldsymbol{r}\, \boldsymbol{j}_i(\boldsymbol{r}, t) \right\} \right|^2 \tag{3}$$

where $\Omega$ is the system volume.

**D. Wavelet transform time-frequency analysis**

The time-frequency analysis is performed by continuous wavelet transform[61]:

$$W(\tau, s) = \left| \frac{1}{\sqrt{s}} \int_{-\infty}^{\infty} dt \left\{ \frac{\partial}{\partial t} \int_\Omega d^3 \boldsymbol{r}\, \boldsymbol{j}_i(\boldsymbol{r}, t) \cdot \Phi^* \left( \frac{t - \tau}{s} \right) \right\} \right|^2 \tag{5}$$

where $\tau, s$ is the translation coefficient and scaling coefficient, respectively. $\Phi^*(t)$ is the father wavelet function, the corresponding mother wavelet function is complex Morlet wavelet, which is given by: $\Phi(t) = 1/\sqrt{\pi B} \cdot \exp(-t^2/B) \cdot \exp(j 2\pi C t)$, where B is the band width, C is the center frequency (chosen by proper parameters) and j is the imaginary unit.

## III. RESULTS and DISCUSSION

**A. Literature reproducibility verification**

First of all, we show the harmonics intensity dependence on laser intensity (Figure 1). The dash line are the fits of the data in perturbative regime[62] (the scaling law $I_n \propto I_{laser}^n$, $I_n$ is the intensity of nth harmonic, $I_{laser}^n$ is the nth power of laser intensity). The deviation between simulation results and fitted lines indicates the HHG mechanisms is under non-perturbative regime. Next, to validate the reliability of our calculation approach, we start to reproduce the experimental results[59] of linear polarization in bulk silicon (Figure 1). The laser intensity is set to $0.2\, TW/cm^2$, weaker than experiments ( $0.7\, TW/cm^2$ ) because of the DFT underestimation of bandgap[57,63]. The photon energy and duration of laser are chosen 0.6eV and 50fs. The main features of HHG obtained from our TDDFT simulation are essentially consistent with the experimental observation and corresponding calculation of other groups[59]. Also, we note the small discrepancies are due to lack of consideration for dephasing factors, macroscopic propagation effects and possible instrument detection error[14,59,64–67].

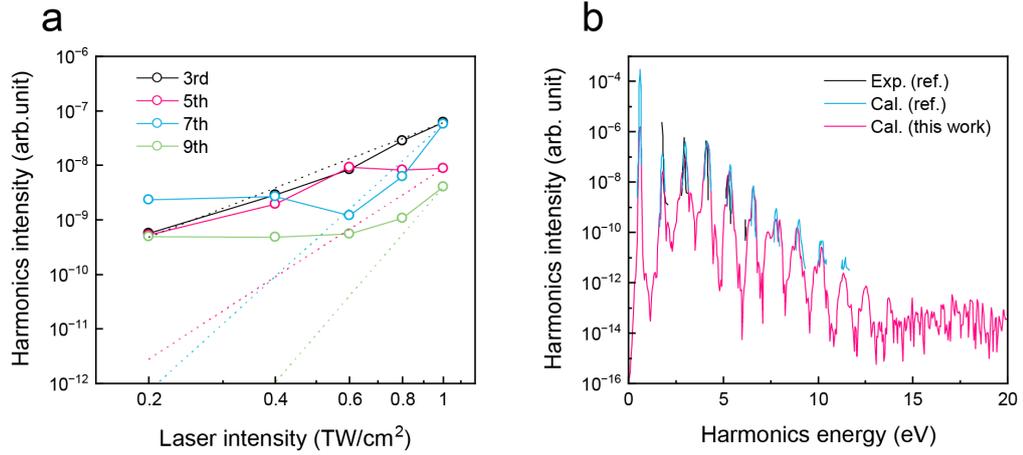

**Figure 1. Experiment v.s. Calculation.** (a) The dash lines show the perturbative harmonic emission, the solid circle-dotted line shows the actual simulation results. (b) The black solid and blue lines are the experimental and calculated results from4 reference[59], respectively, the pink solid line is this work.

### B. Tuning the cutoff energy through bond length modulation

Nowadays, the emission mechanism of solid-state high-order harmonics has been elucidated well[17,19,20], composed by two fundamental processes: (i) intraband process, where nonlinear currents are generated by Bloch oscillations of charge carriers[18,21,64]. (ii) interband process, where electron–hole pairs under a strong laser field could recombine coherently and emit photons[20,22,23,64], and the cutoff energy is determined by the largest band gap encountered at the recombination instant. Notably, the high-frequency cutoff region in solid-state HHG is predominantly domainated by interband processes[14,23,24,64], which provide a clear direction for tuning the cutoff energy. Besides, in solid materials, Decreasement in bond length would enhance the Coulomb repulsion between atomic orbitals, leading to the widening of band gap, and thereby elevating the cutoff energy associated with interband processes. Conversely, increasement in bond length would reduce the bandgap through orbital decoupling, diminishing HHG cutoff energy.

Next, we present the HHG spectra of bulk silicon under various bond length modulations (Figure 2). In the unstrained situation, the cutoff energy is 10.2 eV. When applying a bond length compression of −7.5%, the minimum band gap increases slightly, while the conduction band energies at other k-points rise significantly (Figure 3). As a result, the low-frequency region of the harmonic spectrum shows minimal change in both structure and intensity, but the plateau beyond 10.2 eV is substantially extended to 13.8 eV—an enhancement of nearly one-third. Conversely, under a +7.5% bond length stretching, the minimum band gap is significantly reduced, increasing electron excitation via Landau–Zener tunneling[68,69]. This leads to a notable increase in harmonic intensity in the low-frequency region, whereas the reduced overall band energy shortens the cutoff to approximately 9 eV.

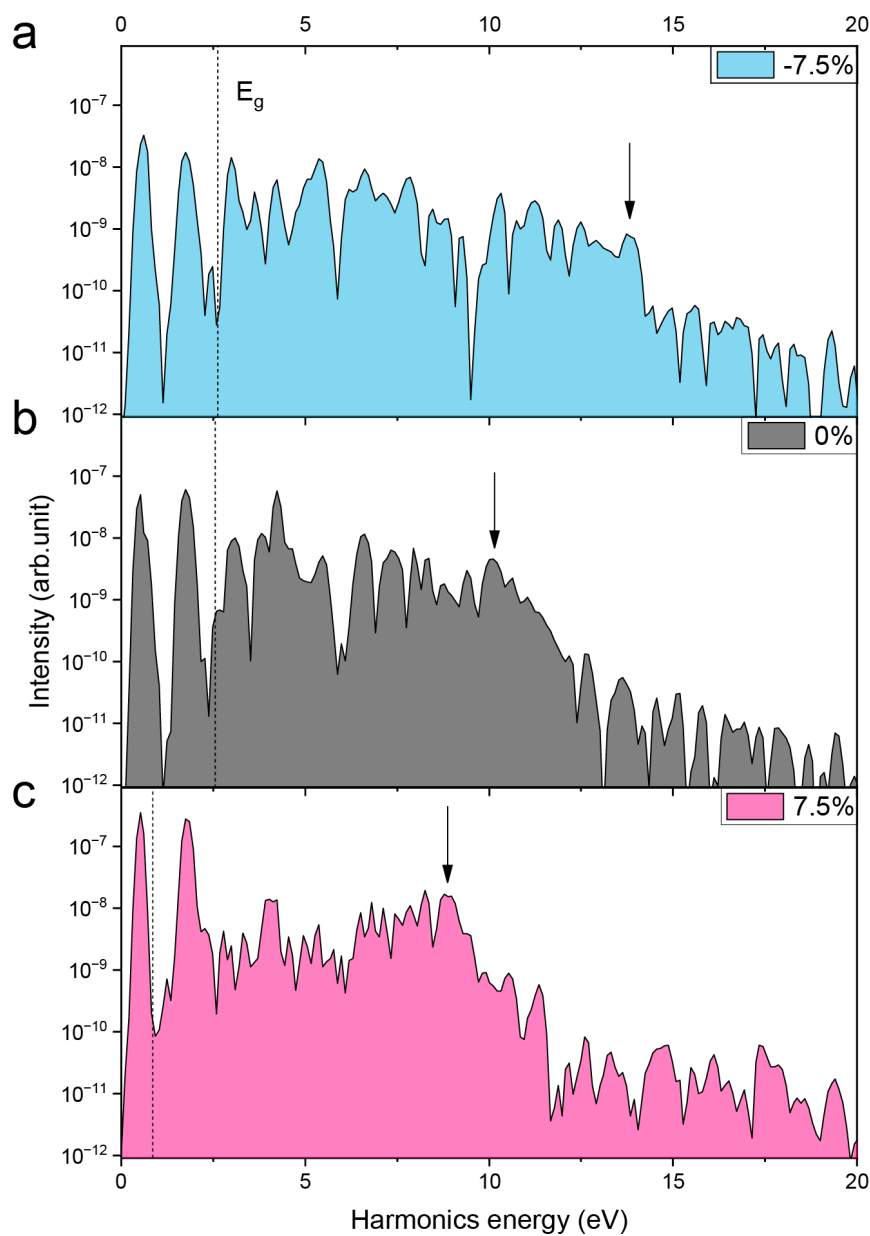

**Figure 2. Cutoff energy under different bond length.** (a), (b) and (c) are the HHG under -7.5%, 0% and 7.5% bond length variation, respectively. The black dotted lines mark the direct minimum band gap. The black arrows mark the corresponding cutoff energy.

Figure 3a, the average intensity of the harmonics increases monotonically with bond length. This is attributed to the reduced band gap, which facilitates greater electron excitation and thus enhances the nonlinear optical response. In addition, the general trend of the cutoff energy also decreases with increasing bond length (Figure 3b). Intriguingly, when the bond length variation exceeds a certain threshold (e.g., ±5%), the cutoff energy saturates or even shows an opposite trend (±10%). This is because excessive bond length modulation leads to drastic changes in the band structure, inducing less electron excitation. As a result,

fewer electrons populate the higher conduction bands (see Supplementary Fig. S2), and the recombination energy of electron–hole pairs cannot increase accordingly. This leads to a decrease in the high-frequency cutoff region, which is dominated by interband processes, even as the bond length is reduced. This phenomenon can be viewed as a competition between harmonic intensity and cutoff energy. While our results show a generally monotonic variation of cutoff energy with bond length, the corresponding harmonic intensity often exhibits an opposite trend[46,47,50]. In actual HHG emission mechanisms, a balance between these two effects typically achieved.

Figure 3b depicts the Coulomb repulsion-modulated bandgap evolution: The momentum-space minimum bandgap energy between conduction band (CBM+3) and valence bands (VBM-2) and corresponding maximum bandgap both decrease monotonically with bond length, where the bandgap energies range covers the cutoff energies corresponding to the different bond lengths. These decreasing trends originate from weakened orbital hybridization at extended atomic distances. With stretching bond length, while interband mechanisms dominants, cutoff energies decrease. Moreover, we visualize the electronic excitations at various bond lengths, revealing that they fully respond to changes in the external electric field. The above conclusion also holds in solid aluminum arsenide, demonstrating that it is material-independent (Fig S3).

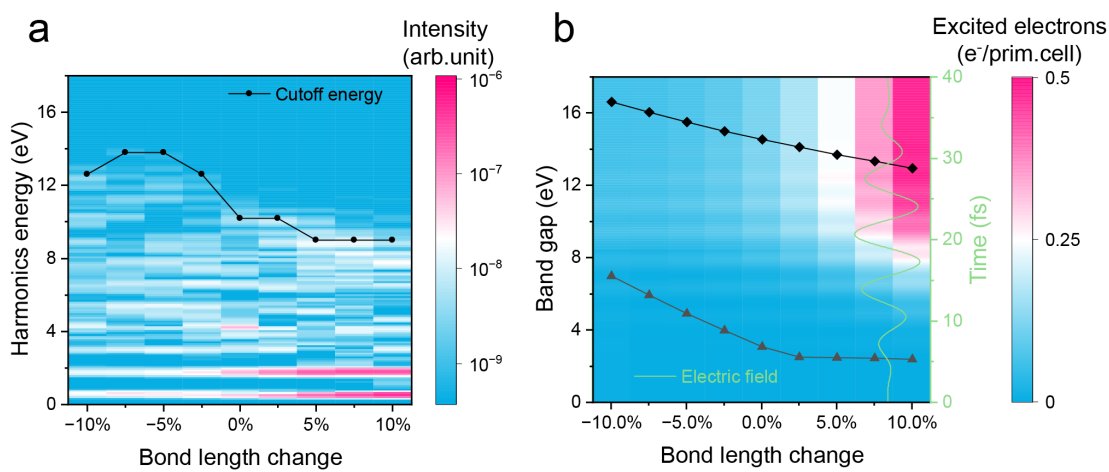

**Figure 3. Bond length engineering of bulk silicon.** (a) The harmonics spectrum in different bond length variance. The circle-dotted lines mark the cutoff energy positions corresponding to different bond length variations. (b) The number of excited electrons under different bond length changes. The green solid line represents the temporal evolution of the electric field. The diamond-dotted and triangle-dotted lines indicate the maximum and minimum band gaps between VBM-2 and CBM+3 along the laser polarization direction **ΓK** corresponding to different bond length variations.

## C. Time-frequency analysis

Figure 4 depicts the time-frequency analyses of HHG emission under three typical bond length variances: -7.5%, 0% (equilibrium), and +7.5%. All spectra exhibit characteristic chirped emission signatures in the high-frequency regime[22,23,64], manifesting as: The positive chirp arises from the short trajectory motion of the electron-hole pair and the negative chirp originates from the long trajectory motion. This spectral fingerprint conclusively confirms interband dominance across bond length variance. In addition, we can also see that with the cutoff energy decreases more obviously increasing bond length, which corresponds to the description mentioned above. (Figure 3a). Notably, we also find the effect of dephasing in the last part of harmonic emission (t > 30fs)[65,67,70,71], which will be aggravated when the bond length increases, which may be related to the failure of coherent recombination of the increased number of electron excitations.

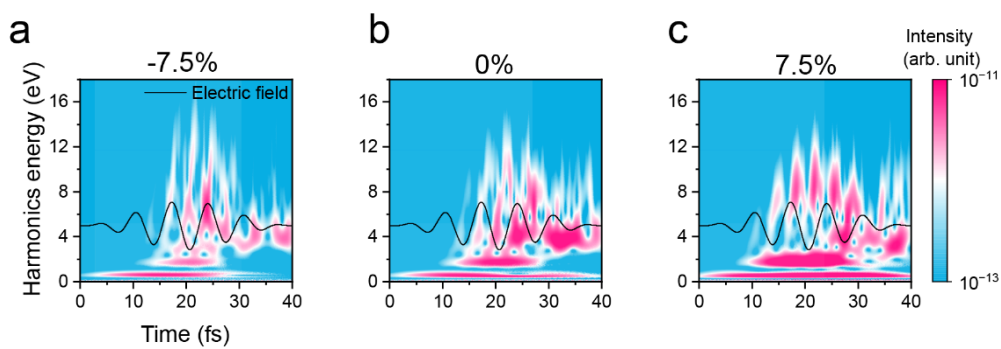

**Figure 4. Wavelet transform of harmonics with bond length in silicon.** (a), (b) and (c) are the wavelet transforms in -7.5%, 0 and 7.5% bond length variance. The black solid line in the figure depicts the electric field.

In addition to the bulk silicon material, we also performed simulations on cubic phase aluminum arsenide (AlAs). Similarly, we observed that the bond length modulates the harmonic cutoff energy (Figure 5). As the bond length increases, the cutoff energy generally decreases. Moreover, in the process of bond length variation, a saturation region of the cutoff energy appears, and even a noticeable decline (up to −10%) can be observed. The average harmonic intensity also decreases with increasing bond length. The results behavior just like the silicon's results, which indicate that the modulation of the cutoff energy by bond length is independent of the specific material. Therefore, by compressing the bond length—experimentally achievable through the application of compressive strain—the harmonic cutoff energy can be significantly enhanced. This provides a promising pathway toward the generation of isolated attosecond pulses[3,24].

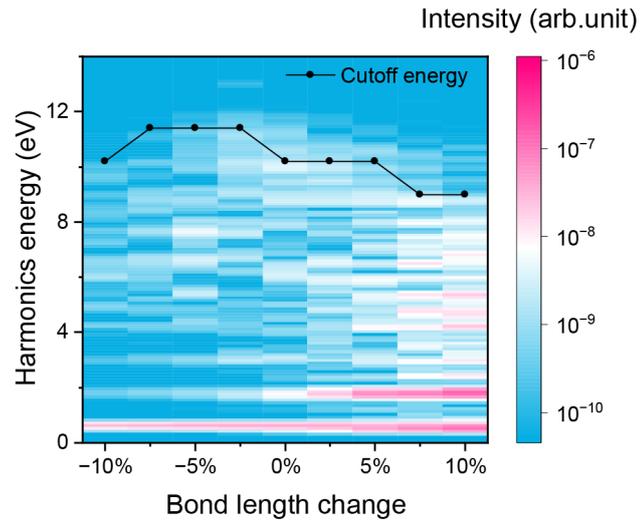

**Figure 5. Bond length engineering of bulk aluminum arsenide.** The harmonics spectrum in different bond length variance. The circle-dotted lines mark the cutoff energy positions corresponding to different bond length variations.

## IV. CONCLUSION

In summary, we theoretically analyzed the modulation of the HHG cutoff energy in solids by varying the bond length. Our first-principles calculations reveal that atomic orbital repulsion modulated by bond-length variations alters bandgap characteristics, thereby modifying interband contributions. As a result, the high-frequency plateau of interband harmonics extends significantly, contributing to an overall increase in the cutoff energy. Further bond length variation, the cutoff energy exhibits a saturation behavior and may even begin to decrease. This non-monotonic trend arises from the competition between the rapidly changing band gap and the corresponding electronic excitation processes. The proposed scheme can be experimentally realized by applying strain, providing an effective strategy to enhance the HHG cutoff energy in solid-state. Future studies will focus on establishing universal scaling laws between cutoff energies and bond lengths in complex materials, a critical step toward engineering optimized HHG performance through targeted material design.

**ACKNOWLEDGEMENTS**

# APPENDIX A: Linear Dependence of Cutoff Energy on Field Strength

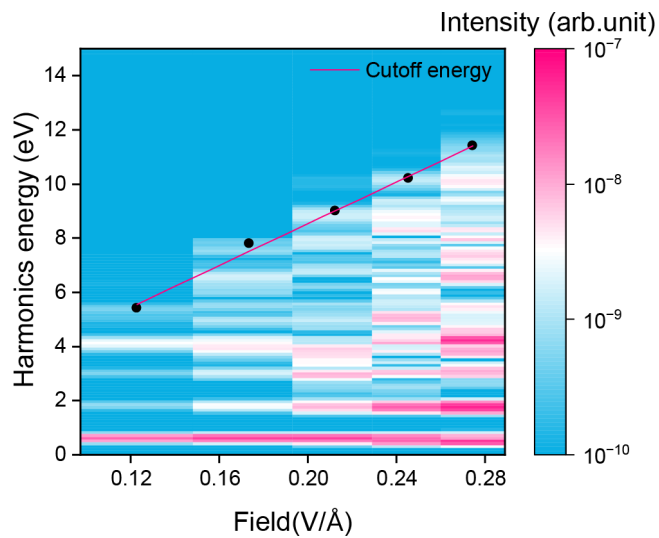

**Figure S 1. The dependence of the cutoff energy on the field strength in silicon.** Black dots denote the cutoff energies obtained at various field strengths, and the red line indicates a linear fit, confirming the proportional scaling behavior in solids.

# APPENDIX B: Main temporal occupation in silicon under different bond length change

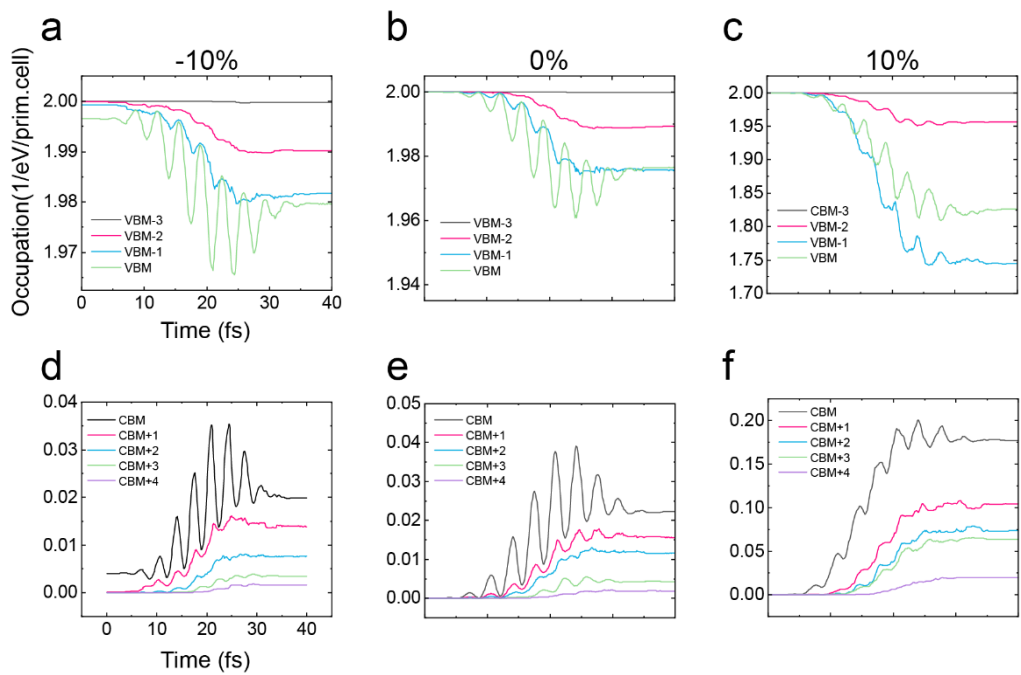

**Figure S 2.** (a), (b) and (c) are the main valance bands electron occupation in excitation under the bond length change of -10%, 0%, 10%, respectively. (d), (e) and (f) are the corresponding main conduction bands electron occupation.

# APPENDIX C: Temporal excited electrons in different bond length change of aluminum arsenide

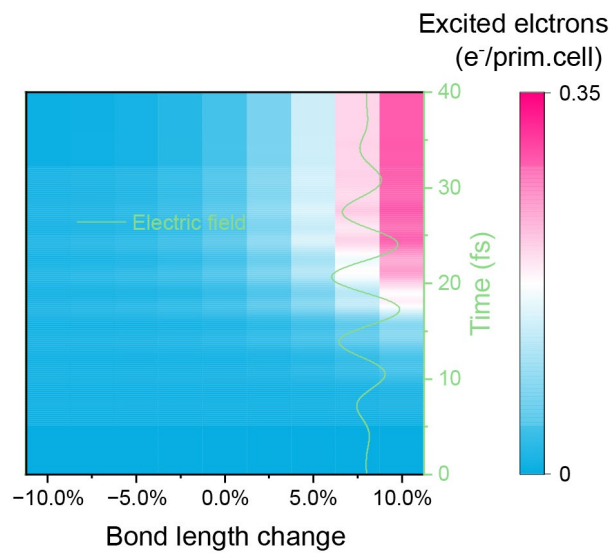

**Figure S 3.** Temporal electron excitation in different bond length change. The solid green line is the electric field.

# APPENDIX D: Bandstructure in different bond length variance

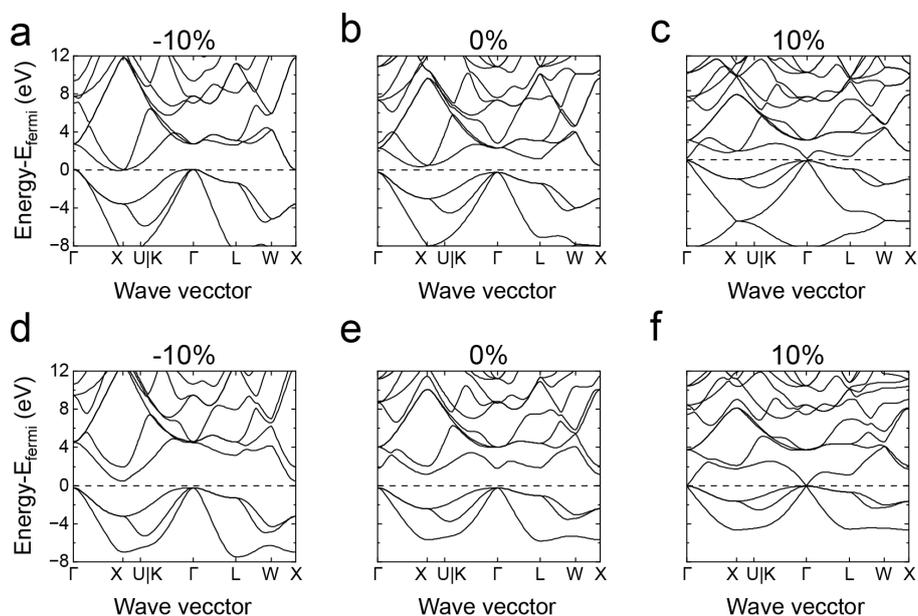

**Figure S 4. Bandstructure comparison between different bond length variance.** (a), (b) and (c) are bandstructure of bulk silicon in -10%, 0 and 10%, respectively. (c), (d) and (e) are bandstructure of bulk aluminum arsenide in -10%, 0 and 10%, respectively.